\documentclass[11pt,a4paper]{article}
\pdfoutput=1
\usepackage{jheppub}
\usepackage{amsmath}
\usepackage{wasysym}
\usepackage{verbatim}
\usepackage{bbm}

\usepackage{appendix}
\usepackage{graphicx}
\usepackage{mathrsfs}
\usepackage{caption}
\usepackage{float}
\usepackage{subfig}
\usepackage{amssymb}
\usepackage{textcomp}
\usepackage{tikz}
\usepackage{xcolor}

\usepackage[T1]{fontenc}
\usepackage[utf8]{inputenc}
\usepackage{amsmath}
\newcommand{\e}{\epsilon}

\newcommand{\p}{\partial}
\newcommand{\refb}[1]{(\ref{#1})}

\renewcommand{\O}{{\mathcal{O}}}

\newcommand{\z}{{\bar z}}

\newcommand{\non}{\nonumber}

\newcommand{\be}[1]{ \begin{equation}\label{#1} }
\newcommand{\ee}{\end{equation}}
\newcommand{\bea}[1]{\begin{eqnarray}\label{#1} }
\newcommand{\eea}{\end{eqnarray}}
\newcommand{\bes}{\begin{subequations}}
\newcommand{\ees}{\end{subequations}}

\renewcommand{\a}{\alpha}

\renewcommand{\t}{\tau}

\newcommand{\D}{\Delta}

\renewcommand{\a}{\alpha}

\renewcommand{\t}{\tau}

\title{3d Carrollian Chern-Simons theory \& 2d Yang-Mills}
\author[a]{Arjun Bagchi,}\author[b]{Arthur Lipstein,}\author[c]{Mangesh Mandlik,}\author[d]{Aditya Mehra.}
\author{\\}

\affiliation[a]{Indian Institute of Technology Kanpur, Kalyanpur, Kanpur 208016. INDIA. \\}
\affiliation[b]{Department of Mathematical Sciences, Durham University, Stockton Road, DH1 3LE,
Durham, United Kingdom.\\}
\affiliation[c]{Department of Physics, Indian Institute of Technology (Indian School of Mines) Dhanbad, Jharkhand 826004, India.\\}
\affiliation[d]{School of Basic and Applied Sciences, JSPM University, Gate No. 720, Wagholi, Pune 412207,
India.\\}
\emailAdd{abagchi@iitk.ac.in}
\emailAdd{arthur.lipstein@durham.ac.uk}
\emailAdd{manmangesh@gmail.com}
\emailAdd{aditya1.mehra@gmail.com}
\abstract{With the goal of building a concrete co-dimension one holographically dual field theory for four dimensional asymptotically flat spacetimes (4d AFS) as a limit of AdS$_4$/CFT$_3$, we begin an investigation of 3d Chern-Simons matter (CSM) theories in the Carroll regime. We perform a Carroll (speed of light $c\to0$) expansion of the relativistic Chern-Simons action coupled to a massless scalar and obtain Carrollian CSM theories, which we show are invariant under the infinite dimensional 3d conformal Carroll or 4d Bondi-van der Burg-Metzner-Sachs (BMS$_4$) symmetries, thus making them putative duals for 4d AFS. Concentrating on the leading-order electric Carroll CSM theory, we perform a null reduction of the 3d theory. Null reduction is a procedure to obtain non-relativistic theories from a higher dimensional relativistic theory. Curiously, null reduction of a Carrollian theory yields a relativistic lower-dimensional theory. We work with $SU(N) \times SU(M)$ CS theory coupled to bi-fundamental matter and show that when $N=M$, we obtain (rather surprisingly) a 2d Euclidean Yang-Mills theory after null reduction. We also comment on the reduction when $N \neq M$ and possible connections of the null-reduced Carroll theory to a candidate 2d Celestial CFT.}

\preprint{}

\begin{document}

\maketitle

\section{Introduction}
The Holographic Principle \cite{tHooft:1993dmi, Susskind:1994vu} has been our preferred path in attempts to understand the quantum nature of gravity in recent years. After the initial ideas originating from the area law of black hole entropy, holography has become almost synonymous with its formulation in Anti de Sitter (AdS) spacetimes through the celebrated AdS/CFT correspondence \cite{Maldacena:1997re}. The Maldacena conjecture  gave us the first concrete dual pair involving type IIB superstring theory on AdS$_5 \times \mathbbm{S}^5$ in the bulk and $\mathcal{N}=4$ SU$(N)$ Supersymmetric Yang-Mills theory on the four-dimensional (4d) flat boundary of AdS$_5$. This is sometimes called the  AdS$_5$/CFT$_4$ correspondence to distinguish it from similar correspondences in other dimensions. 

\medskip

The more recent AdS$_4$/CFT$_3$ correspondence connects type IIA superstring theory on AdS$_4 \times \mathbbm{CP}^3$ with ABJM theory \cite{Aharony:2008ug}, which is a $\mathcal{N}=6$ Superconformal Chern-Simons matter theory with a gauge group $U(N) \times U(N)$ living on the 3d boundary of AdS$_4$. When the string coupling becomes large, type IIA superstring theory goes over to M-theory and hence for generic values of parameters, ABJM theory is dual to M-theory on AdS$_4 \times\mathbbm{S}^7/\mathbbm{Z}_k$. There is also a generalisation of ABJM theory to unequal gauge groups $U(M) \times U(N)$ \cite{Aharony:2008gk}. For more details on the subject, the reader is pointed to the review \cite{Bagger:2012jb}. 

\medskip

Of late, there is a renewed interest in the formulation of holography beyond its original home in AdS, specifically to asymptotically flat spacetimes (AFS). The case of 4d asymptotically flat space in the bulk is of particular interest because of its obvious connection to the real world. There has been a wealth of new connections in the infra-red established between seemingly unrelated corners of asymptotic symmetries, soft theorems and memory effects \cite{Strominger:2017zoo}. Questions of holography in this context have followed in a natural way. 

\medskip

There are now two principle routes to flat holography, viz. Celestial and Carrollian holography. Celestial holography is the proposal that the holographic dual to 4d AFS is a 2d relativistic CFT which lives on the celestial sphere at null infinity. This makes use of the fact that the bulk Lorentz group acts as global conformal transformations on the celestial sphere. The reader is pointed to the recent reviews \cite{Strominger:2017zoo, Pasterski:2021raf, Pasterski:2021rjz, Raclariu:2021zjz} and the references within. Carrollian holography, on the other hand, proposes a co-dimension one hologram in terms of a 3d {\em Carrollian} CFT. A Carrollian theory can be obtained from a relativistic one by sending the speed of light $c$ to zero \cite{LBLL, SenGupta:1966qer} and these are naturally defined on null surfaces. In contrast with Celestial holography, the Carrollian version takes into account the whole Poincare group which now acts as global Carrollian conformal transformations on the whole of the null boundary, crucially keeping track of the null direction. An incomplete set of references on Carrollian holography is \cite{Bagchi:2016bcd, Bagchi:2022emh, Donnay:2022aba, Donnay:2022wvx, Bagchi:2023fbj, Salzer:2023jqv, Saha:2023hsl, Nguyen:2023vfz, Bagchi:2023cen, Mason:2023mti, Alday:2024yyj, Ruzziconi:2024zkr} and older work in this direction, especially in the context of lower dimensions include \cite{Bagchi:2010eg, Bagchi:2012cy,Barnich:2012aw,Bagchi:2012yk,Bagchi:2012xr,Barnich:2012xq, Barnich:2012rz,Bagchi:2014iea,Hartong:2015usd}. 

\medskip

The approaches to flat holography have been principally bottom up, with Celestial holography relying on bulk physics to learn about the features of the dual 2d CFT, and Carrollian holography mainly adopting a similar approach. However see some recent attempts at top-down approaches involving twistor theory \cite{Costello:2022wso,Costello:2022jpg}. It is natural to attempt to build a theory of flat holography by taking a systematic limit of AdS/CFT \cite{Susskind:1998vk,Polchinski:1999ry,Giddings:1999jq} and some recent attempts in this direction include \cite{Ball:2019atb,Casali:2022fro,Bagchi:2023fbj,Bagchi:2023cen,Alday:2024yyj}. We will be interested in this line of inquiry and will focus on 4d AFS. The large radius limit of AdS induces a Carrollian limit in the boundary CFT \cite{Bagchi:2012cy}. With this in mind, our aim is to build the Carrollian equivalent of the ABJM model to connect this to the flat version of the AdS$_4$/CFT$_3$ correspondence. In this paper, we take the first steps towards this broader goal. We construct the Carrollian limit of Chern-Simons (CS) matter theories in $d=3$. 

\medskip

It is by now well known that Carrollian limits come in two varieties called the electric and magnetic limits. Given the action of a relativistic quantum field theory, one can systematically expand out the relevant dynamic fields in powers of the speed of light (this expansion is called the Carroll or $c$-expansion, where $c$ is the speed of light) and the leading term in this action is what goes under the name of the Electric theory. This is, by construction, invariant under Carroll symmetries. The Carrollian electric theories exhibit ultralocal correlation functions containing spatial delta-functions. Such correlators of Carrollian CFTs can be mapped to S-matrix elements in the bulk 4d asymptotically flat spacetimes by the so-called modified Mellin transformation \cite{Bagchi:2022emh, Banerjee:2018gce, Banerjee:2019prz}. Electric Carrollian CFTs are thus prototypical of holograms of flat spacetime. In our paper, we will mostly be interested in Electric Carrollian theories. Magnetic Carrollian theories arise out of the next-to-leading order (NLO) terms in the above mentioned $c$-expansion. The NLO term by itself is not Carroll boost invariant and in order to restore Carrollian symmetries, one needs to put in appropriate Lagrange multipliers. We will briefly look at Magnetic Carrollian CSM theories in two appendices at the end of the paper.     

\medskip

One of the important differences between holography in AdS$_4$ and 4d AFS is the symmetry structure at the boundary. 
A usual recipe for holography is to consider the asymptotic symmetry group (ASG) as the symmetry that dictates the dual field theory. The ASG  is the group of allowed diffeomorphisms for a given set of boundary conditions modded out by the trivial diffeomorphisms. For many cases, as with AdS$_4$, the ASG is simply the isometry group of the background i.e. SO(3,2). In 4d AFS, however, the ASG at its null boundary enhances from the usual Poincare group ISO(3,1) and becomes the infinite dimensional 4d Bondi-van der Burg-Metzner-Sachs (BMS$_4$) group \cite{Bondi:1962px,Sachs:1962zza}. The 3d dual field theory is hence supposed to inherit this infinite dimensional asymptotic BMS$_4$ symmetry from the bulk \cite{Bagchi:2016bcd}. Although this process is non-trivial from the point of view of the Carrollian limit of the CS-matter theory, we will show later in the paper that the 3d Carrollian field theory that we obtain in the limit does admit this infinite dimensional symmetry structure. For the uninitiated, this may seem like a magic trick since the original theory only had finite dimensional symmetries. BMS symmetries are isomorphic to conformal Carroll symmetries which are conformal isometries of the background null structure \cite{Henneaux:1979vn, Duval:2014uoa, Duval:2014uva} and hence the degeneration of the background Lorentzian structure to form the Carrollian structure gives rise to these infinite symmetries. {\footnote{The expectation that the theories obtained in the Carrollian limit would lead to infinite dimensional symmetries in generic dimension was shown to be true at the level of equations of motion for a wide variety of theories in \cite{Bagchi:2019xfx}.} We elaborate on this later in the paper. 

\medskip

The main surprise in our paper comes in the next part of our analysis. In this work, we are develop a specific 3d Carrollian CFT as a putative dual to a gravitational theory in 4d AFS. As we mentioned above, there is also the Celestial approach which proposes a 2d dual relativistic CFT. The 2d Celestial CFT does not depend on the null direction and lives only on the celestial sphere. In an attempt to obtain a 2d Celestial CFT from a 3d Carrollian one, we propose to reduce the 3d Carrollian theory along the null direction. The null reduction of the non-Abelian Carrollian CS matter theory interestingly leads to a 2d Euclidean Yang-Mills theory. The choice of matter here is crucial. We find that only bifundamental matter leads to 2d non-abelian Yang-Mills, while fundamental matter leads to 2d electrodynamics. The Carroll limit of the bosonic version of the ABJM theory with SU(N) $\times$ SU(N) gauge group will lead us to SU(N) Yang-Mills theory in 2d. We also comment on the more general SU(N) $\times$ SU(M) theory. We may expect the null-reduced theory to represent a 2d Celestial CFTs, but a priori Yang-Mills theory in $d=2$ is not conformally invariant. We argue that the theory one gets from the limit inherits scale invariance, and hence full conformal invariance in $d=2$, through the process of null reduction.  
 
\medskip
An outline of the rest of the paper is the following. We take a quick tour of Carrollian and Conformal Carrollian symmetries in Sec.~\ref{symm}. Here we also touch upon aspects of representation theory we would need later in the paper. We focus on Abelian Chern-Simons matter theories in Sec.~\ref{acsm} and explain the $c$-expansion and obtain the Electric and Magnetic Carroll CSM theories. We discuss the emergence of infinite dimensional conformal Carroll symmetries of the Electric theory in the main text while the symmetry structure of the Magnetic sector is discussed in Appendix~\ref{sym-mag}. We then give some details of the null reduction of Carroll CSM theories and obtain 2d electrodynamics starting from the electric theory. The magnetic theory is discussed in Appendix~\ref{nr-mag}. Sec.~\ref{ncsm} contains the generalisation to non-Abelian CSM theories, its Carrollian construction and the details of the null reduced theory which now becomes a 2d SU(N) Yang-Mills if we begin with bi-fundamental matter in CS theory with gauge group SU(N) $\times$ SU(N).  We also outline the construction for the general SU(N) $\times$ SU(M) theory and discuss how the null-reduced theory shows an emergent 2d conformal symmetry making it a candidate 2d Celestial CFT. We conclude with various remarks.

\bigskip

\section{Carroll and Conformal Carroll Symmetries}\label{symm}
Carroll symmetry, first introduced by Levy-Leblond \cite{LBLL} and Sengupta \cite{SenGupta:1966qer}, has become very important of late with emerging applications in a wide variety of physical scenarios, starting from condensed matter \cite{Bidussi:2021nmp,Bagchi:2022eui} and ultra-relativistic fluids \cite{Bagchi:2023ysc,Bagchi:2023rwd} to gravitational physics \cite{Donnay:2019jiz, deBoer:2021jej} and string theory \cite{Bagchi:2013bga,Bagchi:2015nca,Bagchi:2020fpr, Bagchi:2023cfp}. These symmetries arise naturally on null surfaces and hence are found on the event horizons of generic black holes and also at the asymptotic null boundary of flat spacetimes, where the symmetries enhances to their conformal version. The latter is where we would be interested in for our explorations in this paper. In order to set up our calculations in the coming sections, below we give a quick summary of Carroll and conformal Carroll symmetry first from an algebraic and then from a geometric point of view. 

\subsection{Algebraic and Geometric preliminaries}
The Carroll algebra is an In{\"o}n{\"u}-Wigner contraction of the relativistic Poincare algebra where one takes the speed of light to zero ($c\to0$). The conformal Carroll can be obtained by a similar contraction of the relativistic conformal algebra. Starting with the differential representation of the relativistic conformal algebra:
\bea{}
J_{\mu\nu} = x_\mu \p_\nu - x_\mu \p_\nu, \quad P_\mu = \p_\mu, \quad D= x^\mu \p_\mu, \quad K_\mu = 2x_\mu x^\nu \p_\nu-x^\nu x_\nu\p_\mu
\eea
one can take the $c\to0$ limit by sending $t\to \e t, \, x_i\to x_i$ to get the set of generators for the conformal Carroll algebra:
\bes\bea{}
&& H=\p_t, \quad P_i=\p_i, \quad J_{ij}=x_i \p_j-x_j \p_i, \quad B_i=x_i \p_t \\
&& D= t \p_t+x^i \p_i, \quad K= x^i x_i \p_t, \quad K_j=2x_j(t\p_t+x^i\p_i)-(x^i x_i)\p_j.
\eea\ees
The non-zero  commutation relations of these above generators that form the conformal Carrollian algebra are:
\bea{}\label{algebra}
&&\non [J_{ij}, B_k ]=\delta_{k[j}B_{i]}, ~ [J_{ij}, P_k ]=\delta_{k[j}P_{i]},~ [J_{ij}, K_k ]=\delta_{k[j}K_{i]}, ~ [B_i,P_j]=-\delta_{ij}H,\\
&&\non  [B_i,K_j]=\delta_{ij}K,~ [D,K]=K,~[K,P_i]=-2B_i,~[K_i,P_j]=-2D\delta_{ij}-2J_{ij},\\ &&[H,K_i]=2B_i,~[D,H]=-H, ~[D,P_i]=-P_i,~[D,K_i]=K_i.
\eea
The sub-algebra $\{J_{ij}, B_i, P_i, H\}$ forms the Carroll algebra. We will now focus on (2+1)-dimensions. Let us recombine the above generators as
\bes{}\label{dsssa}
\bea{} L_{0}=\frac{1}{2}(D+iJ_{xy}),~~L_{-1}=-\frac{1}{2}(P_x -iP_y),~~L_{1}=\frac{1}{2}(K_x +iK_y), \\
\bar{L}_{0}=\frac{1}{2}(D-iJ_{xy}),~~\bar{L}_{-1}=-\frac{1}{2}(P_x +iP_y),~~\bar{L}_{1}=\frac{1}{2}(K_x -iK_y), \\
M_{00}=P_{0},~~M_{01}=B_x-iB_y,~~M_{10}=B_x+iB_y,~~M_{11}=K_{0}.
\eea\ees
Using the differential representation of the conformal Carroll algebra and the definitions \eqref{dsssa}, we obtain a suggestive form for the Conformal Carroll generators:
\bea{car-gen}&&L_n= z^{n+1}\p_{z} + \frac{1}{2}(n+1) {z^n}t\p_t ,~
\bar{L}_n=\bar{z}^{n+1}\p_{\bar{z}} + \frac{1}{2} (n+1)\bar{z}^nt\p_t,~M_{nm}=z^{n}\bar{z}^{m}\p_t. \quad \eea
where $z=x+iy$ and $\bar{z}=x-iy$. The conformal Carroll algebra now takes the form
\bes{}\label{ds1}
\bea{}&&[L_{n},L_{m}]=(n-m)L_{n+m},~~[\bar{L}_{n},\bar{L}_{m}]=(n-m)\bar{L}_{n+m},\\&&
~[L_{n},M_{rs}]=\Big(\frac{n+1}{2}-r\Big)M_{(n+r)s},~~[\bar{L}_{n},M_{rs}]=\Big(\frac{n+1}{2}-s\Big)M_{r(n+s)}. \\
&&~[M_{rs},M_{pq}] = 0. 
\eea\ees
where $n=-1,0,1$ and $ r,s= 0, 1$. 
If we now extend the generators \eqref{car-gen} for arbitrary integer $n, r, s$, the algebra above  \eqref{ds1} is infinite dimensional. This algebra is isomorphic to the four dimensional Bondi-van der Burg-Metzner-Sachs algebra (BMS$_4$) which is the asymptotic symmetry algebra of asymptotically flat 4d spacetimes at the null boundary \cite{Bondi:1962px, Sachs:1962zza}.

We now give a geometric account of these symmetries \cite{Henneaux:1979vn,Duval:2014uoa, Duval:2014uva}. In flat space, it is very evident the Carroll limit makes the Minkowski metric degenerate. The metric with covariant indices becomes: 
\be{}
\eta_{\mu\nu} = \begin{pmatrix} - c^2&0&0\\ 0 & 1 & 0\\0&0&1 \end{pmatrix}, \quad \eta_{\mu\nu} \xrightarrow{c\to 0} h_{\mu\nu} = \begin{pmatrix} 0&0&0\\ 0 & 1 & 0\\0&0&1 \end{pmatrix}, 
\ee
while the contravariant version takes the following form
\be{}
\eta^{\mu\nu} = \begin{pmatrix}  -\frac{1}{c^2}&0&0\\ 0 & 1 & 0\\0&0&1 \end{pmatrix}, \quad -c^2 \eta^{\mu\nu} \xrightarrow{c\to 0} \Theta^{\mu\nu} = \begin{pmatrix} 1&0&0\\ 0&0&0\\0&0&0 \end{pmatrix} = \theta^\mu \theta^\nu, \, \, \text{where} \, \,  \theta^\mu = \begin{pmatrix} 1\\0\\0  \end{pmatrix}.
\ee
It is clear from the above that we have 
\be{}
h_{\mu\nu}\theta^\nu = 0
\ee
One can generalise this structure to define general Carrollian manifolds with the pair $(h_{\mu\nu}, \theta^\mu)$. Formally, a Carroll manifold is defined as a $d$-dimensional manifold endowed with a degenerate symmetric positive covariant tensor field $h_{\mu\nu}$ and nowhere vanishing vector field $\theta$ which generates the kernel of $h$. This is a ``weak'' Carrollian structure as opposed to a ``strong" structure which also requires the existence of a symmetric affine connection compatible with both $h$ and $\theta$. 

\medskip

\noindent The Carroll algebra is obtained as the isometry of a flat Carroll manifold
\be{}
\mathcal{L}_\zeta \theta^\mu = 0, \quad \mathcal{L}_\zeta h_{\mu\nu} = 0. 
\ee
Here $\mathcal{L}_\zeta$ represents a Lie derivative along the vector field $\zeta$. This actually leads to an infinite dimensional algebra, which reduces to the finite dimensional Carroll algebra we obtained above in the limit when we restrict to linear functions. We shall mostly be interested in the conformal structures on these manifolds. The conformal isometry is generated by
\begin{equation}
	\mathcal{L}_{\zeta}\theta^\mu=\lambda \theta^\mu, \quad \mathcal{L}_{\zeta}h_{\mu\nu}=-2\lambda h_{\mu \nu}.
\end{equation}
Here $\lambda$ is the conformal factor{\footnote{In general, one could choose different conformal factors $\lambda_1$ and $\lambda_2$ for $\theta$ and $h$ and this would lead to the so-called $N$-conformal Carroll algebras, where $N=-\lambda_2/\lambda_1$ and this is related to the anisotropy factor $z=N/2$ which dictates the relative scaling of space and time under dilatations. From the point of view of holography of asymptotically flat spacetimes, where the bulk is a 4d relativistic spacetime, we are interested in 3d field theories that have uniform scaling of space and time, $z=1$ and the above choice is valid.}}. For flat Carroll backgrounds, the solution to the conformal isometry equations above is given by:
\begin{align} \label{kill}
	\xi=\Big(\alpha(x^i)+\frac{t}{2}\partial_if^i(x^j)\Big)\partial_t+f^i(x^j)\partial_i.
\end{align}
Here $x^i$ denotes the $(d-1)$ spatial directions of the $d$-dimensional Carroll manifold. $\alpha(x^i)$ are arbitrary functions of these spatial coordinates and parametrise supertranslations. $f^i(x^j)$ also satisfy conformal killing equations on the spatial slice. We are interested in the case $d=3$ and hence here $f^i(x^j)$ are restricted to be to be holomorphic/anti-holomorphic functions, i.e. $f\equiv f(z)$ and $\bar{f} \equiv \bar{f}(\bar{z})$. It is clear from the above that we can define the generators of the algebra of Carrollian conformal isometry as follows
\be{}
L(f) = f(z)\p_z + \frac{t}{2}\partial_zf(z)\, \partial_t, \quad L(\bar{f}) = \bar{f}(\z)\p_\z + \frac{t}{2}\partial_\z \bar{f}(\z) \partial_t, \quad M(\a) = \a(z,\z)  \partial_t. 
\ee
If we break this up into modes
\bea{}
f(z) &=& \sum_n a_n z^{n+1}, \quad  \bar{f}(\z) = \sum_n {\bar{a}}_n \z^{n+1}, \quad \a(z, \z) = \sum_{r,s} b_{r,s} z^r \z^s\cr
L(f) &=& \sum_n a_n L_n, \quad L(\bar{f}) = \sum_n {\bar{a}}_n \bar{L}_n, \quad M(\a) = \sum_{r,s} b_{r,s} M_{r,s}
\eea
it is straight-forward to check that the generators are the same as \refb{car-gen} and obey the infinite dimensional BMS$_4$ algebra. 

\subsection{Aspects of representation theory}
In this subsection, we briefly recall aspects of representations of Carrollian CFTs. The construction of the representations of Carrollian CFTs is  similar to the relativistic conformal case. Our construction here would be important to understand the symmetries of the specific Carrollian field theories, i.e. the Carroll CSM theories we will focus on later in the paper. 

Let us consider how conformal Carrollian symmetry acts on a generic field $\Phi$ which can be looked upon as a multiplet of different fields $\phi_i$: 
\be{}
\Phi = \begin{pmatrix} \phi_1 \\ \vdots \\ \phi_n \end{pmatrix}.
\ee
We first focus on the little group that keeps the origin ($t=0, x_i=0$) invariant. This is the subgroup generated by the rotations, Carroll boosts, dilatations, and Carroll SCTs. The action of the generators on $\Phi$ is given by 
\bes{}
\bea{}&&
[J_{ij},\Phi(0)]=\mathcal{S}_{ij}\Phi(0),~~[B_{i},\Phi(0)]=\mathcal{B}_i \Phi(0),~~[D,\Phi(0)]=\Delta \Phi(0),\\&&
~[K_i,\Phi(0)]=k_i\Phi(0),~~[K,\Phi(0)]=k\Phi(0).
\eea\ees
The little group generators form a matrix representation at the origin. We can set $k$ and $k_i$ to zero as a consequence of the algebra of these generators, and this is very similar to the usual relativistic CFT analysis. The representations of the whole conformal Carroll algebra are induced from this. The transformations of the fields under the action of the different generators of the algebra at arbitrary points are given by using the translation generators on the generators to move them to act on the field at that point 
\be{op}
\O(t,x_i) = e^{iHt} e^{iP_i x^i} \O(0) e^{-iHt} e^{-iP_i x^i}, 
\ee
where $\O$ represents a generic operator, and in this case a member of the generators, and by repeated use of the Baker-Campbell-Hausdorff formula. This yields:  
\bes{}
\bea{}
&&[J_{ij},\Phi(t,x_i)]= (x_i\p_j - x_j\p_i + \mathcal{S}_{ij}) \Phi(0), \\ && [B_i, \Phi(t, x_i)] = (x_i\p_t + \mathcal{B}_i)\Phi(0),\\
&& [ D, \Phi (t,x_i)] =  (t \partial _t +x^i \partial _i + \Delta) \Phi(0),\\ && [K, \Phi(t, x_i)] = (x^2\p_t + 2x_i\mathcal{B}_i)\Phi(0),\\
&&[K_i,\Phi(t,x_i)]=(2x_i \Delta-2x_j \mathcal{S}_{ij} +2t \mathcal{B}_i +2 t x_i\p_t +2x_ix_j\p_j -x^2 \p_i)\Phi(0).
\eea\ees
In the Carrollian CFTs, we label fields by their dilatation weight $\D$ and consider various spins $\mathcal{S}_{ij}$. The non-trivial features are encoded in the boost matrices $\mathcal{B}_i$, as we will see below. 

\subsection*{Action of infinite dimensional generators on fields in 3d}
We now focus on $d=3$ and discuss aspects of the representations of the infinite dimensional algebra. We define primaries of the whole infinite dimensional conformal Carroll algebra. All fields are labelled under $L_0$ and $\bar{L}_0$
\be{lo}
[L_0, \Phi] = h \Phi, \quad [\bar{L}_0, \Phi] = {\bar{h}} \Phi
\ee
Here $h+{\bar{h}} = \Delta$ and $h-\bar{h} = \mathcal{S}$. Drawing analogies with usual CFT, we define Carrollian primaries are those for which the weights cannot be lowered further:
\be{hw}
[L_n, \Phi] = 0, \quad [\bar{L}_n, \Phi] = 0, [M_{r,s}, \Phi] = 0, \quad \forall n, r, s >0 
\ee
Quasi-primaries are primaries with respect to the global Poincare sub-algebra. In $d=3$, the algebra of the spin matrices related to Carroll boosts $(\mathcal{B}_x,\mathcal{B}_y)$ and rotations $\mathcal{S}$ is given by
\be{JB}
[\mathcal{S},\mathcal{B}_x]=-\mathcal{B}_y,~~[\mathcal{S},\mathcal{B}_y]=\mathcal{B}_x,~~[\mathcal{B}_x,\mathcal{B}_y]=0.
\ee
The commuting nature of the boosts makes it possible to have different boost labels for a particular spin. 

\paragraph{Spin 0 case:}We first look at the scalar representation. This is simply obtained by setting 
\be{} \mathcal{S}=\mathcal{B}_x=\mathcal{B}_y=0.
\ee
With the input above, we can write down the transformation of the primaries $\Phi(t,z,\bar{z})\equiv \phi(t,z,\bar{z})$ for the whole infinite dimensional algebra: 
\bes{}\label{st2}
\bea{}&&[M_{nm},\phi(t,z,\bar{z})]=z^n \bar{z}^m \p_t\phi(t,z,\bar{z}),\\
&&[L_n,\phi(t,z,\bar{z})]=\frac{1}{2}[(z^{n}(n+1)(\D_{\phi}+t\p_t)+2z^{n+1}\p_z)]\phi(t,z,\bar{z}),
\\&&[\bar{L}_n,\phi(t,z,\bar{z})]=\frac{1}{2}[(\bar{z}^{n}(n+1)(\D_{\phi}+t\p_t)+2\bar{z}^{n+1}\p_{\bar{z}})]\phi(t,z,\bar{z}).
\eea\ees
Here, $\D_{\phi}$ denotes the scaling weight of field $\phi$. The subscripts $(n,m)>0$. This is again done by translating the generator to $(x,t)$ by \refb{op} and using the BCH formula. We also invoke \refb{hw}. 
\paragraph{Spin 1 case:}
The spin 1 representation of rotation means that we have 
\be{}
\mathcal{S}_{ij}=\begin{bmatrix}
0 & 0 & 0\\
0 & 0 & -1 \\
0 & 1 & 0
\end{bmatrix}.
\ee
We now have options for our boost generators consistent with \refb{JB}. One of this is the trivial representation: 
\be{triv}
\mathcal{B}_x=\mathcal{B}_y=0.
\ee
The non-Lorentzian nature of the algebra means that one can have more than one representation for the boost generators corresponding to a particular spin. We will be interested in the non-trivial representation:
\bea{boo}
\mathcal{B}_{x}=
\begin{bmatrix}
0 & 0 & 0\\
1 & 0 & 0 \\
0 & 0 & 0
\end{bmatrix},~
\mathcal{B}_{y}=
\begin{bmatrix}
0 & 0 & 0\\
0 & 0 & 0 \\
1 & 0 & 0
\end{bmatrix}.
\eea
These non-trivial boost matrices described are non-diagonalisable that means the components of spinning primaries would mix under boost transformations. We will work in a basis where the spin 1 $\Phi$ field is given by: 
\be{}
\Phi(t,z,\bar{z}) = \begin{pmatrix} a_t (t, z, \z) \\ a_z (t, z, \z) \\ a_\z(t, z, \z) \end{pmatrix}
\ee  
where $a_z=(a_x-ia_y)$ and $a_{\bar{z}}=(a_x+ia_y)$. The action of supertranslation on the different components is given by: 
\bes{}\label{Ma}
\bea{}&&[M_{nm},a_t]=z^n \bar{z}^m \p_ta_t,\\&&
[M_{nm},a_z]=z^n \bar{z}^m \p_t a_z+2nz^{n-1} \bar{z}^{m} a_t,\\&&
[M_{nm},a_{\bar{z}}]=z^n \bar{z}^m \p_t a_{\bar{z}}+2mz^{n} \bar{z}^{m-1} a_t.
\eea\ees
Notice that the Jordan block structure of the boosts mean that $a_t$ is present in the transformation of $a_z, a_\z$ but only transforms into itself under supertranslations. 
Similarly, the action of superrotations on the components are given by
\bes{}\label{La}
\bea{}&&[L_n,a_t]=\left[z^{n+1}\p_z+\frac{1}{2} z^{n}(n+1)(\D+t\p_t)\right]a_t,\\&&
[L_n,a_z]=\left[z^{n+1}\p_z+\frac{1}{2} z^{n}(n+1)(\D+1+t\p_t)\right]a_z+2tn(n+1)a_t z^{n-1},\\&&
[L_n,a_{\bar{z}}]=\left[z^{n+1}\p_z+\frac{1}{2} z^{n}(n+1)(\D-1+t\p_t)\right]a_{\bar{z}}.
\eea\ees
and for the anti-holomorphic counterpart: 
\bes{}\label{bLa}
\bea{}&& [\bar{L}_n,a_t]=\left[\z^{n+1}\p_\z+\frac{1}{2} \z^{n}(n+1)(\D+t\p_t)\right]a_t,,
\\ &&
[\bar{L}_n,a_z]=\left[\z^{n+1}\p_\z+\frac{1}{2} \z^{n}(n+1)(\D-1+t\p_t)\right]a_z,\\&&
[\bar{L}_n,a_{\bar{z}}]=\left[\z^{n+1}\p_\z+\frac{1}{2} \z^{n}(n+1)(\D+1+t\p_t)\right]a_{\bar{z}}+2tn(n+1)a_t\bar{z}^{n-1}, 
\eea\ees
Notice that the different components have different scaling dimensions. Comparing with \refb{lo}, we see that
\be{h}
h_{a_t} = \bar{h}_{a_t} = \frac{\D}{2}; \quad h_{a_z} = \frac{\D+1}{2}, \bar{h}_{a_z} = \frac{\D-1}{2}; \quad h_{a_\z} = \frac{\D-1}{2}, \bar{h}_{a_\z} = \frac{\D+1}{2}.
\ee
One can similarly build higher spin representations. There will be more choices for non-trivial boost matrices as one increases the spin, with all the lower spin boost matrices showing up. For the purposes of this paper, we will be concerned with spin-0 and spin-1 cases. 

\bigskip \bigskip

\section{Abelian Chern-Simons coupled to scalar matter}\label{acsm}
Our goal in this paper is to construct Carrollian versions of Chern-Simons matter theories. We will focus on 3 dimensions. In this section, we give an overview of the basic construction of these theories from the point of view of an expansion in powers of the speed of light $c$ following \cite{deBoer:2021jej} and demonstrate the technique for the Abelian CS theory before we move onto the more interesting non-Abelian case in the next section. This section provides a warm-up for the more involved case to be discussed later. We will also comment on the symmetries of the actions derived and the most interesting point, the dimensional reduction of the Carrollian CSM theory.   

We begin with the well-known relativistic $U(1)$ Chern-Simons theory and couple this to a scalar. This theory is described by the action:
\be{act1}
S=\int dt d^2x \, \left\{ \frac{k}{4\pi} \e^{\mu\nu\rho} A_{\mu}\p_{\nu}A_{\rho} +(D_{\mu}\phi)^{*}(D^{\mu}\phi) \right\} ,\ee
where $\mu=0,1,2$, $k$ is the level of the Chern-Simons term and $D_\mu$ is the gauge covariant derivative: $D_{\mu}=\p_{\mu}-ieA_{\mu}$. 
We note that under a general coordinate transformation, the gauge field transforms as: 
\be{}
\delta A_{\mu}=\xi^{\nu}\p_{\nu}A_{\mu}+\p_{\mu}\xi^{\nu}A_{\nu}.
\ee
We would be interested in splitting up spatial and temporal components in order to consider the Carroll limit. The restriction of the above general coordinate transformation to a Lorentz transformation is 
\be{}
\xi^{0}= \frac{\beta_i x^i}{c}, \quad \xi^{i}=\frac{x^{0}}{c}\beta^{i}=t\beta^{i}
\ee
The gauge field  $A_\mu$ and real scalar field transforms under Lorentz boosts as
\bes
\bea{} 
&& \delta A_{\mu}=ct \beta^i \p_i A_{\mu}+\frac{1}{c} \beta_i x^i \p_t A_{\mu}+\bar{\delta}A_{\mu},~~ \text{where} ~ \bar{\delta}A_{0}=\beta^i A_i, \, \bar{\delta} A_i= \beta_i A_0, \\
&& \delta \phi =ct \beta^i \p_i \phi +\frac{1}{c} \beta_i x^i \p_t \phi.
\eea
\ees

\subsection{Carrollian expansion}
We will now construct the Carrollian version of the CS theory coupled to a scalar field. We will use an expansion of all fields in a power series in $c^2$ \cite{deBoer:2021jej}. The leading term would become what is known as the Electric Carroll theory, while the sub-leading term, with appropriate modifications, becomes the Magnetic theory. The fields in our theory are expanded as: 
\be{exp}
A_{t}=\sum^{\infty}_{n=0}c^{\lambda}c^{2n}a^{(n)}_{t},~A_{i}=\sum^{\infty}_{n=0}c^{\lambda}c^{2n}a^{(n)}_{i},~\phi=\sum^{\infty}_{n=0}c^{\gamma}c^{2n}\phi^{(n)}.
\ee
where we use $A_t =c A_{0}$. We find the transformation rules of the fields at a generic level $(n)$ by considering the expansion of the relativistic fields again. Let us specifically look at the transformation under boosts. We define 
\be{} \beta_i=c b_i, \ee 
where $b_i$ is the Carroll boost parameter. The fields then transforms as
\bes \bea{}
&&\delta a_t^{(n)}=b_i x^i \p_t a_t^{(n)}+t b^i \p_i a_t^{(n-1)}+ b^i a_i^{(n-1)},\\&&
   \delta a_i^{(n)}=b_j x^j \p_t a_i^{(n)}+ b^j t \p_j a_{i}^{(n-1)}+b^i a_t^{(n)}.
\eea\ees 
where for $n=0$, the transformations are included using $a_{\mu}^{(-1)}=0$. It is straight-forward to see that the leading $n=0$ transformations are identical to what we had derived earlier from the representations of the conformal Carroll algebra in \refb{Ma}. In conclusion, the set $(a^{(0)}_t,a^{(0)}_i)$ acts like a vector field with respect to Carroll transformations. These rules are also applicable for the scalar field. The resultant  higher modes in the expansion transforms into each other under these boosts.

\subsection{Electric and Magnetic Actions}
We will now study the expansion of Chern Simon theory coupled to scalar field. The action \eqref{act1} with explicit $c$ factors is given by
\be{act} S=\int dt d^2x \,\frac{k}{4\pi} \Big[\e^{\mu\nu\rho} A_{\mu}\p_{\nu}A_{\rho}\Big]-\frac{1}{c^2}(D_{t}\phi)^{*}(D_{t}\phi)+(D_{i}\phi)^{*}(D_{i}\phi)=\int dt d^2x \, \mathcal{L}.\ee
We will plug \eqref{exp} into \eqref{act} and extract the leading and subleading pieces. We will take $\lambda=\gamma-1$ since we wish to keep the Chern-Simons term at leading order. Interestingly, we get two distinct theories corresponding to $\lambda=0$ and $\lambda \neq 0$. For $\lambda \neq 0$, it is a straight-forward exercise to check that the interaction terms between the gauge fields and scalars (in the covariant derivative) disappear. We will thus focus on the $\lambda=0$ sector alone. 

\medskip

The leading order Lagrangian, which we will call the Electric Carroll Lagrangian is given by: 
\be{el-ac}
\mathcal{L}_0=\frac{k}{4\pi} \e^{\mu\nu\rho} a_{\mu}^{(0)}\p_{\nu}a_{\rho}^{(0)} - (D_{t} \phi)^{(0)*}(D_{t}\phi)^{(0)},
\ee
where we have used the abbreviation $(D_{t}\phi)^{(0)} = (\p_t-iea^{(0)}_{t})\phi^{(0)}$. We will see below that this Lagrangian has Carroll and indeed (infinite) conformal Carroll symmetries. 

\medskip

The next-to-leading order (NLO) Lagrangian is given by: 
\bea{}&&\non\hspace{-1cm}\mathcal{L}_1=\frac{k}{4\pi} \e^{\mu\nu\rho} \Big(a_{\mu}^{(1)}\p_{\nu}a_{\rho}^{(0)}+a_{\mu}^{(0)}\p_{\nu}a_{\rho}^{(1)}\Big)- (D_t\phi)^{(1)*} (D_t \phi)^{(0)}-(D_t \phi)^{(0)*} (D_t\phi)^{(1)} \\&&+(D_i\phi)^{(0)*}(D_i\phi)^{(0)},\eea
where we have defined
\be{}
(D_i \phi)^{(0)} = D^{(0)}_i\phi^{(0)}= (\p_i - ie a^{(0)}_i)\phi^{(0)}, \quad (D_{t}\phi)^{(1)}=\p_t\phi^{(1)}-iea^{(0)}_{t}\phi^{(1)}-iea_t^{(1)}\phi^{(0)}
\ee
This Lagrangian is not Carroll invariant, specifically it is not Carroll boost invariant. In order to rectify this, we modify it by adding Lagrange multipliers $(\chi_\mu, \xi)$ to make it Carroll boost invariant. We re-write the Lagrangian after adding Lagrange multipliers, to get: 
\bea{}
\non\mathcal{L}_1=&& \frac{k}{4\pi} \e^{\mu\nu\rho} \Big(\tilde{\chi}_{\mu} \p_{\nu}a_{\rho}^{(0)}+a_{\mu}^{(0)}\p_{\nu}\tilde{\chi}_{\rho}\Big) - (\tilde{\xi}^* +ie\tilde{\chi}_t \phi^{*(0)}) (D_t\phi)^{(0)} +(D_t\phi)^{(0)*} (\tilde{\xi}-ie\tilde{\chi}_t \phi^{(0)}) \\
&& \qquad +(D_i\phi)^{(0)*}(D_i\phi)^{(0)},
\eea
where we have redefined $\tilde{\xi}=(D_t^{(0)}{\phi}^{(1)}+\xi)$ and $\tilde{\chi}_{\mu}=(a_{\mu}^{(1)}+\chi_{\mu})$. As we elaborate in Appendix A, by ascribing certain transformation properties to the Lagrange multipliers, the above Lagrangian can be made Carroll invariant. In conclusion, the Carrollian Chern-Simons matter theories that we would be interested in have the following Lagrangians:
\bea{}
\text{Electric:}&& \quad \mathcal{L}_{e}=\frac{k}{4\pi} \e^{\mu\nu\rho} a_{\mu}\p_{\nu}a_{\rho} - (D_{t} \phi)^*(D_{t}\phi), \label{car-el}\\
\text{Magnetic:}&& \quad \mathcal{L}_m=\frac{k}{4\pi} \e^{\mu\nu\rho} \Big({\chi}_{\mu} \p_{\nu}a_{\rho}+a_{\mu} \p_{\nu} {\chi}_{\rho}\Big)-\Big[J_t^* (D_t\phi)+(D_t\phi)^* J_t\Big]+(D_i\phi)^{*}(D_i\phi) \nonumber \\
\label{car-mag}\eea
Here we have $D_{t}=\p_t-iea_{t}, D_{i}=\p_i-iea_{i}$, $J_t\equiv \xi-ie \chi_t \phi$, $a_{t}^{(0)}\equiv a_{t}$ and $\tilde{\chi}\equiv \chi$ and so on. We have dropped all superscripts on the fields.  

\subsection{Symmetries for the electric action}
We now briefly delve into the symmetries of the electric action \refb{car-el}. The transformation of the vector fields $\Phi=(a_t, a_z, a_\z)$ under the conformal Carroll algebra are given by the equations \refb{Ma}, \refb{La} and \refb{bLa}. The transformation of the scalar $\phi$ is given by \refb{st2}. The dilation weights of the different components of the vector field are given by
\be{Da}
\D_{a_\mu} = 1 \Rightarrow h_{a_t} = \bar{h}_{a_t} = \frac{1}{2}; \quad h_{a_z} =1, \, \bar{h}_{a_z} = 0; \quad h_{a_\z} = 0, \, \bar{h}_{a_\z} = 1.
\ee
For the scalar we have
\be{Dp}
\D_\phi = \frac{1}{2}.
\ee
These can be deduced from the relativistic theory in the same way as we constructed the change under the boosts. The dilatation weights don't change under the limit $c\to$ since the dilatation generator $D= t\p_t+x^i\p_i$ does not change under contraction. 

We can now explicitly check for the invariance of the Lagrangian  \refb{car-el} under supertranslations the action of which on the fields are given by \refb{Ma} and \refb{st2}. This yields
\be{} 
\delta_{M}\mathcal{L}_{e}=\p_t[z^n  \bar{z}^m \mathcal{L}_{e}].
\ee
So we see that the electric action is invariant under infinite dimensional supertranslations. Similarly, the action of the ``holomorphic'' superrotations are given by \refb{La} and \refb{st2}. This gives
\be{} 
\delta_{L}\mathcal{L}_{e}=\p_t\Big[\frac{1}{2}z^n(n+1)t\mathcal{L}_{e}\Big]+\p_z\Big[z^{n+1}\mathcal{L}_{e}\Big].
\ee
In the above, we have explicitly used the weights \refb{Da} and \refb{Dp}. We thus have only total derivative terms under the variation of the electric Lagrangian and hence the action is invariant under the infinite dimensional superrotations as well. 

Let us put in perspective what we have found. The relativistic Chern-Simons action in 3d coupled to massless scalar matter is conformally invariant, but this symmetry is finite dimensional. We have taken a Carrollian expansion on this action and considered the leading electric Carroll action. This action is now invariant under an {\em infinite dimensional} symmetry, viz. the 3d conformal Carrollian or BMS$_4$ algebra. This theory is a potential model of a field theoretic dual to a gravitational theory in 4d asymptotically flat spacetimes. 

One can also look at the symmetries of the magnetic action \refb{car-mag} and conclude the emergence of infinite dimensional symmetries there. We give the details of this in Appendix~\ref{sym-mag}. 

\subsection{Null reduction of Carrollian theory}
One of the objectives of our work is to relate the two approaches to holography in asymptotically flat spacetimes, the Carroll and the Celestial. As indicated in the introduction, Carrollian holography proposes a co-dimension one dual to 4d asymptotically flat spacetimes living on the entire null boundary, while Celestial holography advocates a co-dimension two dual that resides on the celestial sphere. The 3d Carrollian field theory is defined on the null line $ \mathbbm{R}_u$ as well as the sphere $ \mathbbm{S}^2$ at $\mathscr{I}^\pm$. It is thus natural to ask what happens if we reduce the 3d theory along the null direction and this is what we will do below. 

Before proceeding, it is important to remind the reader that when one does a null reduction of a relativistic theory in $(d+1)$ dimensions, one ends up with a Galilean theory in $d$ dimensions. In order to null reduce, the relativistic theory is written in lightcone coordinates $x^\pm =\frac{1}{\sqrt{2}}(x^0\pm x^d)$ and then the derivative along $x^+$ is set to zero: $\p_+ =0$. For the purposes of this quick comment, let us focus on 4d theories. In terms of the metric, in the lightcone coordinates in four dimensions, we have
\be{}
 \eta_{4\times 4} =  \left[
    \begin{array}{cccc}
    0 & 1 & 0 & 0 \\
      \cline{2-4}  \multicolumn{1}{c|}{1} & {0} & 0 & \multicolumn{1}{c|} {0} \\
      \multicolumn{1}{c|}{0} & 0 & 1 & \multicolumn{1}{c|} {0} \\ 
      \multicolumn{1}{c|}{0} & 0 & 0 & \multicolumn{1}{c|} {1} \\ 
      \cline{2-4}
    \end{array}
    \right] = \begin{pmatrix} 0 & \t_{3\times1} \\ \t_{1\times 3} & \, \, h_{3\times3} \end{pmatrix}
\ee
The null reduction focuses on the lower $3\times 3$ block. This is a degenerate metric giving rise to a 3d Galilean structure. Now let us attempt the same on a 4d Carrollian theory. We know that here we already have a degenerate metric: 
\be{}
g_{4\times 4} =  \left[
    \begin{array}{cccc}
    0 & 0 & 0 & 0 \\
      \cline{2-4}  \multicolumn{1}{c|}{0} & {1} & 0 & \multicolumn{1}{c|} {0} \\
      \multicolumn{1}{c|}{0} & 0 & 1 & \multicolumn{1}{c|} {0} \\ 
      \multicolumn{1}{c|}{0} & 0 & 0 & \multicolumn{1}{c|} {1} \\ 
      \cline{2-4}
    \end{array}
    \right] = \begin{pmatrix} 0 &  0_{3\times1} \\ 0_{1\times 3} & \, \, \delta_{3\times3} \end{pmatrix}
\ee
The null reduction again will focus on the lower $3\times 3$ block, but now in contrast to the relativistic case, we have a 3d Euclidean non-degenerate metric. We might expect that a null reduction of a Carrollian theory thus would generate a Euclidean theory in one lower dimension{\footnote{There has been recent work relating lower dimensional non-Lorentzian theories (both Galilean and Carrollian) to relativistic theories in lightcone coordinates in one higher dimension from a geometric perspective \cite{Bagchi:2024epw}. It would be of interest to see if something similar can be attempted for higher dimensional Carroll theories and lower dimensional relativistic ones.}}. This expectation is borne out by our analyses in this paper. 

\bigskip

Armed with this intuition, we will now Kaluza Klein reduce the Carrollian theory along the null or $t$-direction.  Splitting the space and time indices, we see that the electric Lagrangian is given by
\be{lkj1}
\mathcal{L}_{e}=\frac{k}{4\pi}\e^{txy}[a_t f_{xy} -a_x (\p_t a_y-\p_y a_t) +a_y (\p_t a_x-\p_x a_t) ]-(D_t\phi)^{*}(D_t \phi).
\ee
The process of null reduction, as just mentioned above, means that any derivative in $t$-direction is set to zero. Doing this we get: 
\be{lkj2}
\mathcal{L}_{null-red}=\frac{k}{4\pi}\e^{txy}[a_t f_{xy} +a_x \p_y a_t-a_y \p_x a_t]-e^2a_t^2\phi^* \phi = \frac{k}{2\pi}\e^{txy}a_t f_{xy}-e^2a_t^2\phi^* \phi,
\ee
where we have dropped a total derivative in the intermediate steps. Our aim now is to integrate out the $a_t$ field. The equation of motion of $a_t$ is given by: 
\be{}
\frac{k}{4\pi}\e^{txy} f_{xy}=e^2a_t\phi^* \phi.
\ee
Completing square(s), \eqref{lkj2} can be written as
\bea{}\mathcal{L}_{null-red}=-e^2\phi^*\phi\left(a_t - \frac{k}{4\pi e^2}\frac{f_{xy}}{\phi^*\phi}\right)^2+\left(\frac{k}{4\pi e}\right)^2\frac{f_{xy}^2}{\phi^*\phi}\eea
Classically, the $a_t$ equation of motion suggests that the bracket of the first term vanishes. In the path integral language, the bracket gives a gaussian integral in shifted $a_t$, which just yields a determinant. In either case, we are left with only the second term after integrating out $a_t$. So we find that
\be{}\mathcal{L}_{null-red}= \left(\frac{k}{4\pi e}\right)^2\frac{f_{xy}^2}{\phi^*\phi}
\ee
This is a 2D Euclidean pure Maxwell theory with coupling $\frac{1}{g^2} = \left(\frac{k}{4\pi e|\phi|}\right)^2$, provided $|\phi|$ acquires a vacuum expectation value by some mechanism.

The magnetic Carroll theory can also be null reduced and we provide some details in Appendix~\ref{nr-mag}. This is more involved and we will not be concerned with this in the main body of the paper. 

In conclusion, we have shown that starting with a 3d relativistic Abelian Chern-Simons theory coupled to scalar matter, one can do a Carroll expansion in powers of the speed of light to obtain two Carroll Chern-Simons matter theories in 3d, which exhibit infinite dimensional Conformal Carroll symmetry. Now, null reducing the electric Carroll CS theory and integrating out the $a_t$ field, we have ended up in a 2d Euclidean Maxwell theory. This section provides a warm-up for the more involved non-Abelian case we would be addressing in the coming sections. It is rather curious that one can end up with a lower dimensional Maxwell theory from Chern-Simons theory in this way. 

We started out this sub-section saying that we wanted to relate 3d Conformal Carroll theories to 2d Celestial CFTs via null reductions. We have obtained a 2d Euclidean Maxwell theory. Now Maxwell theory is only classically conformally invariant in $d=4$. So a priori, it is not clear at all that we have ended up with a relativistic CFT in $d=2$. We will however argue that this is the case when we move to the details of the non-Abelian theory in the coming sections.

\bigskip \bigskip

\section{Bifundamental CSM Theories}\label{ncsm}

We will now consider a non-abelian generalisation of our construction in the previous section, viz. a Chern-Simons matter theory with bifundamental matter and gauge group $SU(N)\times SU(M)$. Such theories famously arise in the context of AdS$_4$/CFT$_3$ duality \cite{Aharony:2008ug}. Note that the ABJM theory has $U(N)\times U(N)$ gauge group, the $U(1)\times U(1)$ part of which can be gauge-fixed to a discrete subgroup. We will avoid this subtlety by working with special unitary groups. For simplicity, we will also neglect fermions and scalar potential terms. First we will take the Carrollian limit to obtain a Chern-Simons matter theory with Carrollian conformal symmetry (or BMS$_4$ symmetry), which can be thought of as living at null infinity of Minkowski space, providing a toy model for flat space holography. Then we will perform dimensional reduction along the null direction to obtain a relativistic two-dimensional theory. It is notable if start with a relativistic theory and apply a Carrollian limit followed by a null reduction, we end up with a relativistic theory in one lower dimension. In a sense, we can think of the non-relativistic limit encoded by the null reduction as cancelling out the ultra-relativistic limit encoded by the Carrollian limit. We expect this to be a more general phenomenon. Moreover, we will show that the resulting theory has relativistic 2d conformal symmetry and may therefore be a celestial CFT.

We will show below that upon giving the scalar fields a vacuum expectation value, the null-reduced 3d theory becomes a Euclidean 2d Yang-Mills theory. To our knowledge, such a connection between 3d CSM and 2d YM (YM) theory has not previously been observed. In particular, if $M \leq N$ the gauge group of 2d YM will be $SU(M)$. From this we see that having fundamental matter in 3d (which corresponds to having $M=1$) will lead to an abelian theory in 2d even if the 3d theory has a non-abelian gauge group. Hence, it is crucial to have bifundamental matter in 3d in order to get an interacting theory in 2d. It is intriguing that the necessity of bifundamental matter was previously discovered using completely different reasoning in the context of AdS/CFT \cite{Schwarz:2004yj,Bagger:2007vi,Aharony:2008ug}. This suggests that if a concrete realisation of flat space holography exists, it should indeed arise from taking the flat space limit of AdS/CFT. 

\subsection{Carrollian bifundamental CSM}
We begin by considering the relativistic CS theory coupled to bifundamental scalar matter: 
\begin{align}
S=\int dt\,dx\,dy & \left\{ \frac{ik_{M}}{8\pi}e^{\mu\nu\rho}\operatorname{Tr}_{N}\left(A_{\mu}\partial_{\nu}A_{\rho}+\frac{2i}{3}A_{\mu}A_{\nu}A_{\rho}\right)\right. \nonumber\\
& +\frac{i k_{M}}{8 \pi} \epsilon^{\mu \nu \rho} \operatorname{Tr}_{M}\left(B_{\mu} \partial_{\nu} B_{\rho}+\frac{2 i}{3} B_{\mu} B_{\nu} B_{\rho}\right) \nonumber \\
& \left.+\operatorname{Tr}_{M}\left[\left(D_{\mu} \phi\right)^{\dagger}\left(D_{\mu} \phi\right)\right]\right\}
\end{align}
where $\phi$ is a scalar field in the in $(N, \bar{M})$ representation of $\operatorname{SU}(N) \times \operatorname{SU}(M)$, $A_{\mu}$ and $B_{\mu}$ are $SU(N)$ and $SU(M)$ gauge fields, respectively, and
\begin{equation}
D_{\mu} \phi=\partial_{\mu} \phi-i A_{\mu} \phi+i \phi B_{\mu}.
\end{equation}
We will choose the Chern-Simons levels to be $k_{N}=-k_{M}=k$. We will see later that this choice gives 2d YM theory after taking the Carrollian limit followed by dimensional reduction. It is also the choice which appears in the ABJ(M) theory \cite{Aharony:2008ug,Aharony:2008gk}.

We employ the same Carroll expansion \refb{exp}, but now for both gauge fields $A_\mu$ and $B_\mu$, along with the scalar $\phi$. This results in a generalisation of the Abelian Carroll actions we wrote down earlier. We will focus solely on the leading electric action in this case (but the magnetic case can be similarly obtained). The Carrollian electric non-Abelian CSM action is given by 
\begin{align}
S_{e}=\int d t d x d y & \left\{\frac{i k}{8 \pi} \epsilon^{\mu \nu \rho} \operatorname{Tr}_{N}\left(a_{\mu} \partial_{\nu} a_{\rho}+\frac{2 i}{3} a_{\mu} a_{\nu} a_{\rho}\right)\right. \nonumber \\
& -\frac{i k}{8 \pi} \epsilon^{\mu \nu \rho} \operatorname{Tr}_{M}\left(b_{\mu} \partial_{\nu} b_{\rho}+\frac{2 i}{3} b_{\mu} b_{\nu} b_{\rho}\right) \nonumber \\
& \left.-\operatorname{Tr}_{M}\left[\left(D_{t} \phi\right)^{\dagger} \cdot D_{t} \phi\right]\right\},
\end{align}
where we have
\begin{equation}
D_{t} \phi=\partial_{t} \phi-i a_{t} \phi+i \phi b_{t}.
\end{equation}
This theory can be shown to have infinite dimensional Carrollian conformal symmetry, like its Abelian counterpart, and can be thought of as a CFT living in null boundary of Minkowski space, presumably dual to some gravitational theory in the bulk. 

\subsection{Dimensional Reduction and emergence of 2d Yang-Mills}
In continuation of the construction in the Abelian case, we will now dimensionally reduce along the null direction, $t$. We remind the reader again that this is a null reduction, which normally gives a lower-dimensional non-relativistic theory when applied to a relativistic theory.  However applied to a Carrollian theory, this yields a relativistic Euclidean theory, so in a sense the non-relativistic nature of the null reduction counters the ultra-relativistic nature of the Carroll theory leading to a relativistic theory at the end of the process. When applied to our Carrollian CSM, the lower dimensional theory is again relativistic. We will show that is contains 2d Yang-Mills theory and enjoys 2d relativistic conformal symmetry, and can therefore be interpreted as a celestial CFT. 

To perform the dimensional reduction, simply take $\partial_{t} \rightarrow 0$. After doing so, we obtain
\begin{align}
S_{2 d}=\int d x d y & \left\{\frac{i k}{4 \pi} \operatorname{Tr}_{N}\left(a F_{x y}\right)-\frac{i k}{4 \pi} \operatorname{Tr}_{M}\left(b \tilde{F}_{x y}\right)\right. \nonumber \\
& \left.+\operatorname{Tr}_{M}\left[(a \phi-\phi b)^{\dagger}(a \phi-\phi b)\right]\right\},
\label{2dtheory}
\end{align}
where $a=a_t$, $b=b_t$, and  
\bes
\begin{align}
F_{x y}=&\partial_{x} a_{y}-\partial_{y} a_{x}+i\left[a_{x}, a_{y}\right],\\
\tilde{F}_{x y}=&\partial_{x} b_{y}-\partial_{y} b_{x}+i\left[b_{x}, b_{y}\right].
\end{align}
\ees
We will now integrate out $a, b$. To simplify the analysis and the physical interpretation of the result, we will give $\phi$ a vacuum expectation value (vev). The simplest case is $N=M$. In this case we obtain $SU(N)$ YM. For $M<N$, we get $SU(M)$ YM plus additional terms whose physical interpretation we will discuss later. 

\paragraph{Case 1: $M=N$.} Let us first consider $N=M$. In this case we can set  $\phi=v \mathbbm{1}_{N\times N}$ giving
\begin{align}
S_{2 d}=\int d x d y & \left\{\frac{i k}{8 \pi} \operatorname{Tr}_{N}\left[a_{+}\left(F_{x y}-\tilde{F}_{x y}\right)\right]\right. \nonumber  \\
& \left.+\frac{i k}{8 \pi} \operatorname{Tr}_{N}\left[a_{-}\left(F_{x y}+\tilde{F}_{x y}\right)\right]+v^{2} \operatorname{Tr}_{N}\left(a_{-}^{2}\right)\right\},
\end{align}
where $a_{ \pm}=a+b$. We then find the following equations of motion:
\bes
\begin{align}
& a_{+} \text { eom: } F_{x y}=\tilde{F}_{x y} \\
& a_{-} \text { eom: } a_{-}=-\frac{i k}{8 \pi v^{2}} F_{x y}. 
\end{align}
\ees
Plugging these back into the action then gives
\begin{equation}
S_{2 d}=\frac{1}{g_{y M}^{2}} \int d x d y \operatorname{Tr}_{N}\left(F_{x y}^{2}\right),\,\,\,  g_{\text{YM}}^{2}=\frac{64 \pi^{2} v^{2}}{k^{2}}.
\end{equation}

\paragraph{Case 2: $M<N$.} We now focus on the more complicated case $M<N$. In this case, we may set
\begin{equation}
\phi=v\binom{\mathbbm{1}_{M \times M}}{0_{(N-M) \times M}}.
\end{equation}
It is also convenient to split the gauge fields and field strengths into blocks as follows:
\begin{equation}
a=\left(\begin{array}{ll}
a_{M\times M} & a_{M\times(N-M)}^{\prime\dagger}\\
a_{(N-M)\times M}^{\prime} & a_{(N-M)\times(N-M)}^{\prime\prime}
\end{array}\right),\,\,\,b=b_{M\times M},
\end{equation}
\begin{equation}
F_{xy}=\left(\begin{array}{ll}
F_{xy}^{M\times M} & F_{xy}^{\prime\dagger M\times(N-M)}\\
F_{xy}^{\prime(N-M)\times M} & F_{xy}^{\prime\prime(N-M)\times(N-M)}
\end{array}\right),\,\,\,\tilde{F}_{xy}=\tilde{F}_{xy}^{M\times M}.
\end{equation}
After doing so, we find that
\begin{align}
S_{2 d}=\int d x d y &\left\{\frac{i k}{8 \pi} \operatorname{Tr}_{M}\left[a_{+}\left(F_{x y}-\tilde{F}_{x y}\right)^{M \times M}\right]\right. 
+\frac{i k}{8 \pi} \operatorname{Tr}_{M}\left[a_{-}\left(F_{x y}+\tilde{F}_{x y}\right)^{M \times M}\right] \nonumber \\
& +\frac{ik}{4\pi}\left[\operatorname{Tr}_{N-M}\left(a^{\prime\prime}F_{xy}^{\prime\prime}\right)+\operatorname{Tr}_{N-M}\left(a^{\prime}F_{xy}^{\prime\dagger}\right)+\operatorname{Tr}_{M}\left(a^{\prime\dagger}F_{xy}^{\prime}\right)\right] \nonumber \\
& \left.+v^{2}\left[\operatorname{Tr}_{M}a_{-}^{2}+\operatorname{Tr}_{M}\left(a^{\prime\dagger}a^{\prime}\right)\right]\right\},
\end{align}
where $a_{\pm}=\left(a\pm b\right)_{M\times M}$. We then find the following equations of motion:
\bes
\begin{align}
& a_{+}\text{ eom: }F_{xy}^{M\times M}=\tilde{F}_{xy}^{M\times M} \\
& a_{-}\text{ eom: }a_{-}=-\frac{ik}{8\pi^{2}v^{2}}F_{xy}^{M\times M}\\
&a'\text{ eom: }a'=-\frac{ik}{4\pi v^{2}}F'_{xy}\\
& a^{\prime \prime} \text { eom: } F_{x y}^{\prime \prime}=0.
\end{align}
\ees
Plugging these back into the action finally gives
\begin{equation}
S_{2d}=\frac{1}{g_{\mathrm{YM}}^{2}}\int dxdy\left[\operatorname{Tr}_{M}\left(F_{xy}^{M\times M}\right)^{2}+4\operatorname{Tr}_{M}\left(F_{xy}^{\prime\dagger}F_{xy}^{\prime}\right)\right], \quad \text{where} \, \, g_{\mathrm{YM}}^{2}=\frac{64 \pi^{2} v^{2}}{k^{2}}.
\end{equation}
Note that the first term describes 2d $SU(M)$ YM, while the second term involves the field strength $F^{\prime}_{xy}$ which is an $M\times (N-M)$ non-hermitian matrix. The physical interpretation of the second term is unclear in general, but when $M=1$, $F^{\prime}_{xy}$ is an $(N-1)$-component vector, i.e. $F_{xy}^{\prime}=\left(F_{xy}^{(1)},\ldots,F_{xy}^{(N-1)}\right)$ and the second term reduces to a sum over $(N-1)$ abelian non-Hermitian fields:
\begin{equation}
\operatorname{Tr}_{M}\left(F_{x y}^{\prime+} F_{x y}^{\prime}\right)=\sum_{\alpha=1}^{N-1}\left|F_{x y}^{(\alpha)}\right|^{2}.
\end{equation}
Note that $M=1$ corresponds to having fundamental matter coupled to $SU(N)$ Chern-Simons theory in the original 3d theory but after dimensional reduction we end up with an Abelian theory if even the original theory was non-Abelian. 

From our findings above, we clearly see that having bifundamental matter in three dimensions is required in order to have an interacting theory after dimensional reduction.  Interestingly, the same conclusion was reached from a very different perspective when constructing a consistent example of the AdS$_4$/CFT$_3$ correspondence. We believe that this is not a coincidence. 

\subsection{Hints of 2d relativistic conformal symmetry}

In this sub-section, we will indicate how the 2d theory in \eqref{2dtheory} exhibits an emergent conformal symmetry arising from dimensional reduction. To motivate this, first recall the vector representation of the 3d Carrollian conformal group \refb{car-gen}, which we re-write here for ease of reading: 
\bes
\begin{align}
& L_{n}=z^{n+1} \partial_{z}+\frac{1}{2}(n+1) z^{n} t \partial_{t}, \\
& L_{n}=\bar{z}^{n+1}\partial_\z+\frac{1}{2}(n+1) \bar{z}^{n} t \partial_{t}, \\
& M_{n, s}=z^{r} \bar{z}^{s} \partial_{t}
\end{align}
\ees
Here the first two lines represents the superrotations which close to two copies of Virasoro algebra, but are in an unusual 3d representation with $(t, z, \z)$ and the third line represents the generators of angle-dependent supertranslations along the null direction $t$. Dimensional reduction along the null direction sets the $t$ derivatives to zero, i.e. $\partial_t \equiv 0$ and we are left with 
\begin{equation}
L_{n}=z^{n+1} \partial_{z}, \quad L_{n}=\bar{z}^{n+1} \partial_{z}.
\end{equation}
These are the usual representation of the generators of the two copies of the Virasoro algebra in $d=2$. We thus expect the 2d theory, which is a null-reduced 3d Carrollian CFT,  to have 2d relativistic conformal symmetry. 

Let us now understand how the 2d Yang-Mills theory can have an emergent scale invariance. Looking at the first line in \eqref{2dtheory}, we see that $a,b$ must have scaling dimension zero since the strengths $F_{x y}$ and $\tilde{F}_{x y}$ have scaling dimension two. Applying this to the second line in \eqref{2dtheory} then implies that $\phi$ has scaling dimension one. After giving $\phi$ a vev and integrating out $a,b$ we see that the resulting 2d YM theory is also scale-invariant since $g^2_{\mathrm{YM}}$ has scaling dimension two. In summary, we find that
\begin{equation}
\Delta_{a}=\Delta_{b}=0,\,\,\,\Delta_{\phi}=1.
\end{equation}
The crucial point here is that the fields that are to be integrated out from the 3d theory $a_t = a$ and $b_t=b$ have changed scaling dimensions from what we started out with, as has the field which acquires a vev, i.e. $\phi$. Since $a, b$ are scalars in the 2d picture, it is natural to set the scaling dimension of $a=b=0$.

Although we don't claim to understand the process of null reduction at the level of the representation theory completely, let us attempt some more explanations. We wish to figure out how the 2d conformal representations appear naturally from the 3d Carroll representations under this process. In particular, the transformation of the fields $a_t, a_i$ and $\phi$ in the 3d action before the null reduction are given by Eqs.~\refb{Ma}--\refb{bLa} and \refb{st2}. The process of null reduction would change the dilation weights of $a_t$ and $\phi$. In particular, due to the different scaling dimensions for $a_t$ and $a_i$, the 3d Carroll boosts do not mix these components of the spin-one field into each other. So these objects under Carroll boosts would transform in the trivial representation \refb{triv} instead of the non-trivial one \refb{boo} for the spin-one multiplet. In particular, the transformation of each field would be according to \refb{st2} and doing the null reduction by setting the $t$-derivatives here to zero gives us a natural 2d conformal transformation: 
\bes
\bea{}
&&[L_n,\Phi(z, \z)]=\left[z^{n+1}\p_z +  z^{n}(n+1)h \right]\Phi(z,\bar{z}),\\
&&[\bar{L}_n,\Phi(z,\bar{z})]=\left[\z^{n+1}\p_\z +  \z^{n}(n+1)\bar{h} \right]\Phi(z,\bar{z}),
\eea\ees
where $\Phi(z,\z)= (a_z, a_\z)$. The weights of the fields are give by
\be{}
h_{a_z} = 1, ~ \bar{h}_{a_z}= 0~; \quad h_{a_\z} = 0, ~ \bar{h}_{a_\z}= 1.
\ee 
These follow directly from \refb{h} since $\D_{a_z}= \D_{a_\z}=1$, which does not change with the dimensional reduction. The above transformation can also obtained from Eqs.~\refb{La} and \refb{bLa} by setting $\p_t\equiv0$ and $a_t\equiv0$. It is now straightforward to show that the theory in \eqref{2dtheory} enjoys 2d conformal symmetry.

It is interesting to note that we obtain a theory with relativistic conformal symmetry by performing a null reduction of a theory with Carrollian conformal symmetry. We believe that this mechanism is not special to Carrollian CSM theory, and should hold for any theory which arises from taking the Carrollian limit of a relativistic theory essentially because the non-relativistic limit encoded by the null reduction cancels out the ultrarelativstic limit of the Carrollian limit. 

\bigskip \bigskip

\section{Conclusions}

\subsection{Summary}
Motivated by the ABJM construction of a concrete dual to AdS$_4$ spacetimes in terms of 3d CSM theories, in this paper we laid out the basic construction of a holographic dual to 4d AFS in terms of a 3d Carrollian CSM theory. We arrived at the Carrollian theories by considering a $c$-expansion of the fields in the relativistic theory and showed that the leading Electric Carroll CSM theory has an infinite dimensional BMS$_4$ symmetry. This makes the theory a candidate for a field theory dual to 4d AFS, since it inherits the asymptotic symmetries of the bulk gravitational theory. In Appendix~\ref{sym-mag}, we discuss aspects of the sub-leading magnetic theory, which also exhibits similar symmetry structures. 

We then performed a null reduction of the 3d Carrollian theories. Reducing along the null direction, we ended up with 2d (Euclidean) relativistic theories. The theory we reduced to depended very crucially on the matter content of the parent theory. We considered bi-fundamental matter and non-Abelian relativistic CS theories and then the process of first taking the Carroll limit followed by a null reduction landed us on a 2d Yang-Mills theory with $SU(N)$ gauge symmetry, if we started out with two equal gauge groups $SU(N) \times SU(N)$. For the $SU(N) \times SU(M)$ case ($N>M$), the results were more involved, with a 2d SU(M) YM theory with additional interactions. For fundamental matter, the theory reduced to 2d electrodynamics. This rather surprising connection between 3d CSM theories and 2d YM theories, to the best of our knowledge, is completely novel and could be the tip of the iceberg of a deep connection between 3d-2d theories via this curious ultrarelativistic-nonrelativistic reduction. 

We ultimately provided some hints as to how the 2d YM theory we obtained has an emergent 2d relativistic conformal symmetry and thus may provide a bridge between 3d Carrollian CFTs and 2d Celestial CFTs. We provide more comments below.

\bigskip

\subsection{Discussions and future directions}
Our work raises several tantalising questions and below we discuss some of them. 

\begin{itemize}
\item[$\star$] {\bf{Relating Carroll and Celestial CFTs through null reductions.}} \\
As described in the introduction, in recent years, there has been a major theoretical effort to formulate flat space holography in terms of a 2d CFT living on the sphere at null infinity, known as the Celestial CFT \cite{Strominger:2017zoo,Pasterski:2021raf}.  Given that the 2d theory we obtain by performing a null reduction of a 3d Carrollian CFT has 2d conformal symmmetry, we believe that this theory may provide a concrete relisation of a celestial CFT, or at least be closely related to one. Let us suggest a speculative holographic argument which lends support to this claim. First recall the formula for a bulk-to-boundary propagator for a field dual to a scalar operator with dimension $\Delta$ in a Carrollian CFT \cite{Bagchi:2023fbj}:
\begin{equation}
\tilde{G}_{\triangle}=\frac{1}{(t+q \cdot x)^{\Delta}},
\end{equation}
where $\vec{q}$ is a null vector which can be interpreted as the momentum of a massless particle in 4d Minkowski space. This propagator was derived by writing the AdS$_4$ propagator in 5d embedding corrdinates and taking the flat space limit. If we restrict our attention to one edge of null infinity parametrised by $0<u<\infty$ and impose appropriate boundary conditions, we can extract the zero mode of the operators along this interval of the boundary by simply performing an integral over $u$ as follows:  
\begin{align}
\int_{0}^{\infty} d u G_{\Delta} & =\int_{0}^{\infty} \frac{d u}{(u+q \cdot x)^{\Delta}}=\left.\frac{1}{1-\Delta} \frac{1}{(u+q \cdot x)^{\Delta-1}}\right|_{0} ^{\infty} \\
& =\frac{1}{\Delta-1} \frac{1}{(q\cdot x)^{\Delta-1}},\,\,\, \Delta \neq 1.
\end{align}
We recognise the second line as the bulk-to-boundary propagator in AdS$_3$ which can be derived from the Mellin transform of a plane wave in 4d Minkowski space \cite{Cheung:2016iub}. More generally, performing this Mellin transform maps scattering amplitudes to Celestial correlators \cite{Pasterski:2017kqt}. Hence, dimensional reduction maps a Carrollian CFT operator with scaling dimension $\Delta$  to a celestial CFT operator with scaling dimension $\Delta-1$.

There have been other similar suggestions for relating 3d Carrollian and 2d Celestial CFTs (see e.g. \cite{Donnay:2022wvx}). We hope to follow up on this, specifically in the context of 3d CSM theories we have discussed above. It would also be interesting to explore if there is any relation between 2d YM and other recent proposals for Celestial CFTs \cite{Costello:2023hmi,Stieberger:2023fju,Melton:2024gyu}.  

\item[$\star$] {\bf{Limits and reductions}} \\
We have performed a Carroll limit followed by a null reduction on the 3d relativistic CSM theories to end up with 2d Yang-Mills theories. It would be intriguing to figure out what happens if one does the opposite, i.e. null-reduce the 3d relativistic theory and perform a Carroll limit on the resulting theory and to generalise this story to other spacetime dimensions. We hope to report on this in the near future. 

\item[$\star$] {\bf{Computing correlation functions}} \\
Given a concrete proposal for a Carrollian CFT, it would be of great interest to compute its correlation functions in order to probe the dynamics of the bulk theory. For this puropse, it would be useful to adapt the Feynman rules recently derived for Carrollian YM theories in \cite{Islam:2023rnc} to Carrollian CSM theories. A natural target to derive from the boundary perspective would be tree-level Einstein gravity amplitudes, which were recently mapped to Carrollian correlators in \cite{Bagchi:2023cen,Mason:2023mti}. In general, we expect boundary correlators to produce amplitudes of Einstein gravity plus an infinite tower of higher derivative corrections which arise from the low energy expansion of a UV finite theory of quantum gravity such as string theory. While reproducing bulk locality at four-points may require performing a non-perturbative calculation in the boundary \cite{Maldacena:2015iua}, we should already be able to get some insight into the bulk dynamics by computing three-point functions. Indeed, conformal Ward identities imply that three-point stress tensor correlators in relativistic CFT's must be a linear combination of two different structures which correspond to two-derivative and six-derivative gravitational interactions in the bulk \cite{Osborn:1993cr,Bzowski:2017poo,Farrow:2018yni}, so one expects to have a similar statement for 3d Carrollian CFT's.

\item[$\star$] {\bf{Supersymmetrization}} \\
One of the most important directions is to generalise our discussion to include supersymmetry and in particular figure out what the Carroll limit of 3d $\mathcal{N}=6$ Supersymmetric CS theory is so that we can actually focus on the flat limit of the AdS$_4$/CFT$_3$ correspondence. Supersymmetric versions of Carrollian theories in dimensions higher than two have been addressed in \cite{Bagchi:2022owq}. It would be of interest to use these algebraic structures in the construction of an explicit supersymmetric CSM model. Understanding the analogue of this limit for type IIA string theory on AdS$_4 \times$ CP$^3$ is also an important project, but one may have to work a lot harder for a full string theoretic understanding of the bulk.

\end{itemize}
We hope to come back to these, and other questions of interest, very soon. 

\bigskip

\subsection*{Acknowledgements}
We thank Rudranil Basu, Prateksh Dhivakar, Sudipta Dutta, Romain Ruzziconi, and Akshay Yelleshpur Srikant for useful discussions. AB is partially supported by a Swarnajayanti Fellowship from the Science and Engineering Research Board (SERB) under grant SB/SJF/2019-20/08 and also by SERB grant CRG/2022/006165. AB thanks the participants and organisers of the workshop ``Carrollian Physics and Holography'' organised at the Erwin Schr{\"o}dinger Institute (ESI), University of Vienna, for interesting discussions, and the ESI for hospitality during the visit. AL is supported by an STFC Consolidated Grant
ST/T000708/1.

\appendix
\section*{APPENDICES}
\section{Symmetries of Magnetic limit}\label{sym-mag}
In this appendix, we will look into the magnetic limit and the symmetries of the action. The action in the magnetic limit is given by
\begin{align}\label{rcl}
\mathcal{L}_{mag}=\frac{k}{4\pi}\,\Big[\e^{tij} \Big(\chi_{t} \p_{i}a_{j}-\chi_{i} \p_{t}a_{j}+\chi_{i} \p_{j}a_{t}+a_{t}\p_{i}\chi_{j}-a_{i}\p_{t}\chi_{j}+a_{i}\p_{j}\chi_{t}\Big)\Big] \nonumber \\ -\Big[J_t^* (D_t\phi)+(D_t\phi)^* J_t\Big]+(D_i\phi)^{*}(D_i\phi),
\end{align}
where we have $D_{t}=(\p_t-iea_{t}), D_{i}=(\p_i-iea_{i})$, $J_t\equiv(\xi-ie \chi_t \phi)$, $a_{t}^{(0)}\equiv a_{t}$ and $\tilde{\chi}\equiv \chi$ and so on. 
The equations of motion are
 \bea{}&&
 \frac{k}{2\pi}\e^{tij}\tilde{f}_{jt}+ie[\phi^{*}(D_i \phi)-\phi(D_i \phi)^{*}]=0,~\frac{k}{4\pi}\e^{tij}\tilde{f}_{ij}+ie[\phi J_t^*-\phi^* J_t]=0,\\&&
\frac{k}{4\pi}\e^{tij}f_{ij}-ie[\phi^{*}(D_t \phi)-\phi(D_t \phi)^{*}]=0,~\frac{k}{2\pi}\e^{tij}f_{ti}=0,\\&&
D_t(J_t)-ie\chi_t (D_t\phi)-D_iD_i\phi=0,~ D_t\phi=0. 
\eea
where $\tilde{f}_{ab}=(\p_a\chi_b-\p_b\chi_a)$ and $a=(t,i)$. We will now look at the symmetries of the Lagrangian \eqref{rcl}. 
\paragraph{Boost transformation:} The transformations of the various fields in the Lagrangian under the action of Carroll boosts is given by
\bea{} [B_{i},a_t (x^i,t)] = x_{i}\partial_{t}a_t,\quad [B_{i},a_{j} (x^i,t)] = x_{i}\partial_{t}a_{j} +  \delta_{ij}a_t\\
~[B_{i},\chi_t (x^i,t)] = x_{i}\partial_{t}\chi_t,\quad [B_{i},\chi_{j} (x^i,t)] = x_{i}\partial_{t}\chi_{j} + \delta_{ij}\chi_t,\\
~[B_{i},\phi (x^i,t)] = x_{i}\partial_{t}\phi,\quad [B_{i},\xi (x^i,t)] = x_{i}\partial_{t}\xi +(D_i \phi).
\eea
The boost transformations of the Lagrange multipliers $(\chi_a, \xi)$ are chosen in a manner so as to make sure that the action is invariant under Carroll boosts. Below we see this explicitly. The variation of Lagrangian under boost transformation is given by
\bea{}\delta_{B}\mathcal{L}_{mag}=x_l \p_t \mathcal{L}_{mag}=\p_t[x_l \mathcal{L}_{mag}].\eea
The magnetic action thus is invariant under Carroll boosts. 
\paragraph{Scale transformation:}
The transformation of the fields under dilatations is given by: 
\bea{}[ D, \Phi (x^i,t)] =  (t \partial _t +x^i \partial _i + \Delta_{\Phi}) \Phi, \eea
where $\Phi\equiv (a_t,a_i,\phi,\chi,\xi)$ and $\Delta_{\Phi}$ denotes each fields respected scaling weight.
Using it to understand  the variation of the Lagrangian, we get
\bea{}&&\hspace{-1cm}\delta_D\mathcal{L}_{mag}=\p_l[x^l\mathcal{L}_{mag}]+\p_t[t\mathcal{L}_{mag}]+(2\D_{\phi}-1)[(D_i\phi)^{*}(D_i\phi)]\non\\&&\non+(\D_{\chi}-1)\frac{k}{4\pi}\,\Big[\e^{tij} \Big(\chi_{t} \p_{i}a_{j}-\chi_{i} \p_{t}a_{j}+\chi_{\mu} \p_{\nu}a_{\rho}+a_{t}\p_{i}\chi_{j}-a_{i}\p_{t}\chi_{j}+a_{i}\p_{j}\chi_{t}\Big)\Big]\\&&-\Big(\Delta_{\phi}-\frac{1}{2}\Big)\Big[(\xi^* +ie \chi_t \phi^*) (D_t\phi)+(D_t\phi)^* (\xi-ie \chi_t \phi)\Big]\non\\&&-\Big(\D_{\xi}-\frac{3}{2}\Big)\Big[\xi^* (D_t\phi)+(D_t\phi)^* \xi\Big]-\Big(\D_{\chi}+\D_{\phi}-\frac{3}{2}\Big)ie\chi_t\Big[ \phi^* (D_t\phi)-(D_t\phi)^* \phi\Big]\eea
We have already taken $\D=1$ for the gauge field $(a_t, a_i)$ in the intermediate steps. All extra terms vanishes when we take
\be{sc}\Big[\D=1,\D_{\chi}=1,\D_{\xi}=\frac{3}{2},\D_{\phi}=\frac{1}{2}\Big].\ee
Finally the result becomes
\be{} \delta_D\mathcal{L}_{mag}=\p_l[x^l\mathcal{L}_{mag}]+\p_t[t\mathcal{L}_{mag}].\ee
The magnetic action is thus invariant under scale transformation given the scaling dimensions of the fields \refb{sc}. 
\paragraph{Supertranslation transformation:}
We will now look into the supertranslations and the invariance of the magnetic limit. The transformations of the fields under supertranslation is given by
\begin{itemize}
\item For the scalar $\phi$: \refb{st2}. 
\item For the vector field $\vec{a}=(a_t, a_z, a_\z)$, and Lagrange multiplier $\vec{\chi}= (\chi_t, \chi_z, \chi_\z)$: \refb{Ma}.
\item For the Lagrange multiplier $\xi$: 
\be{} [M_{nm},\xi]=z^n  \bar{z}^m \p_t\xi+nz^{n-1} \bar{z}^m D_z\phi+mz^{n} \bar{z}^{m-1}D_{\bar{z}}\phi.
\ee
\end{itemize}
The variation of \eqref{rcl} under supertranslation comes out to be
\bea{} \delta_{M}\mathcal{L}_{mag}=\p_t[z^n \bar{z}^m \mathcal{L}_{mag}]\eea
The Magnetic Carrollian CSM theory thus has infinite dimensional supertranslation invariance. 
\paragraph{Superrotations transformation:}
We now move on to superrotations. The transformations of the fields under superrotations are given by: 
\begin{itemize}
\item For the scalar $\phi$: \refb{st2}. 
\item For the vector field $\vec{a}=(a_t, a_z, a_\z)$, and the vector Lagrange multiplier $\vec{\chi}= (\chi_t, \chi_z, \chi_\z)$: \refb{La} and \refb{bLa}.
\item For the Lagrange multiplier $\xi$: 
\be{} [L_n,\xi]=\frac{1}{2}[(z^{n}(n+1)(\D_{\xi}+t\p_t)+2z^{n+1}\p_z)\xi+tn(n+1)(D_z\phi)z^{n-1}].
\ee
\end{itemize}
Using the above, the variation under superrotations of \eqref{rcl} comes out to be
\bea{} \delta_{L}\mathcal{L}_{mag}=\p_t\Big[\frac{1}{2}z^n(n+1)t\mathcal{L}_{mag}\Big]+\p_z\Big[z^{n+1}\mathcal{L}_{mag}\Big].\eea
We thus see that the magnetic action is invariant under infinite dimensional superrotations. The magnetic Carrollian CSM action thus has all the infinite dimensional symmetries of the extended BMS$_4$ algebra. 

\section{Null reduction of magnetic theory}\label{nr-mag}
In this appendix, we provide some details of the null reduction of the magnetic Carrollian CSM theory. For simplicity, we focus on the Abelian case. 
The Lagrangian is given by
\bea{}&&\non\hspace{-1.2cm}\mathcal{L}_{mag}=\frac{k}{4\pi}\,\Big[\e^{txy} \Big(\chi_{t} f_{xy}-\chi_{x}f_{ty} +\chi_{y} f_{tx}+a_{t}\tilde{f}_{xy}-a_{x}\tilde{f}_{ty}+a_{y}\tilde{f}_{tx}\Big)\Big]\\&&-\Big[J_t^* (D_t\phi)+(D_t\phi)^* J_t\Big]+(D_x\phi)^{*}(D_x\phi)+(D_y\phi)^{*}(D_y\phi).\eea
In order to null reduce the theory, we set the derivatives $\p_t\equiv0$. The action of the reduced theory thus becomes
\bea{}&&\non\hspace{-1.2cm}\mathcal{L}_{mag}=\frac{k}{4\pi}\,\Big[\e^{txy} \Big(\chi_{t} f_{xy}+\chi_{x}\p_y a_t -\chi_{y} \p_x a_{t}+a_{t} \tilde{f}_{xy}+a_{x}\p_y \chi_t-a_{y}\p_x \chi_t\Big)\Big]\\&&+ie a_t\Big[J_t^* \phi-\phi^* J_t\Big]+(D_x\phi)^{*}(D_x\phi)+(D_y\phi)^{*}(D_y\phi),\eea
taking the total derivatives, we get
\bea{}&&\non\hspace{-1.2cm}\mathcal{L}_{mag}=\frac{k}{2\pi}\,\Big[\e^{txy} \Big(\chi_{t} f_{xy}+a_{t} \tilde{f}_{xy}\Big)\Big]+ie a_t\Big[J_t^* \phi-\phi^* J_t\Big]+(D_x\phi)^{*}(D_x\phi)+(D_y\phi)^{*}(D_y\phi).\eea
The equations of motion for $a_t$ and $\chi_t$ are given by
\bea{}\frac{k}{2\pi}\e^{txy}\tilde{f}_{xy}=-ie[J^*_t \phi-J_t\phi^*],~\frac{k}{2\pi}\e^{txy} f_{xy}-2e^2a_t\phi^*\phi=0.\eea
Looking above, we get two cases. Let us discuss each case in details. First one is integrating out $a_t$ and auxiliary fields and second one is if $\xi$ is not integrated out. 
\subsection*{Integrating out $a_t$ and auxiliary fields}
The magnetic action for an Abelian Chern-Simons field minimally coupled to a scalar, after taking away $\partial_t$ is 
\bea{}&&\non\hspace{-1.2cm}\mathcal{L}_{mag}=\frac{\kappa}{2\pi}\,\Big[\e^{txy} \Big(\chi_{t} f_{xy}+a_{t} \tilde{f}_{xy}\Big)\Big]+ie a_t\Big[J_t^* \phi-\phi^* J_t\Big]+(D_x\phi)^{*}(D_x\phi)+(D_y\phi)^{*}(D_y\phi)\eea
Substituting the definition $J_t = \xi - i\chi_t\phi$, we get
\begin{equation}\label{Magseg}
\mathcal{L}_{mag} = a_t\left(ie\phi\xi^*-ie\phi^*\xi-2e^2\phi^*\phi\chi_t\right)+\frac{\kappa}{2\pi}\left(f_{xy}\chi_t+\tilde{f}_{xy}a_t\right)+|D_i\phi|^2
\end{equation}
The goal is to integrate out $a_t$, $\chi_t$ and $\xi$. A slight generalization of the well known fact that a product can be written as a difference of two squares helps us write the ``quadratic" terms in \eqref{Magseg} as
$$a_t\left(ie\phi\xi^*-ie\phi^*\xi-2e^2\phi^*\phi\chi_t\right) = |\left(a_t+\frac{ie\phi^*}2 \xi-\frac{e^2\phi^*\phi}2\chi_t\right)|^2-|\left(a_t-\frac{ie\phi^*}2 \xi+\frac{e^2\phi^*\phi}2\chi_t\right)|^2$$
Let's define
$$V_1 = a_t+\frac{ie\phi^*}2 \xi-\frac{e^2\phi^*\phi}2\chi_t,~~~~V_2 = a_t-\frac{ie\phi^*}2 \xi+\frac{e^2\phi^*\phi}2\chi_t$$
So the quadratic piece is $V_1^*V_1-V_2^*V_2$. Now let's look at the linear piece. $a_t$ is simply $\frac{V_1+V_2}2$, while $\chi_t = \frac{V_2-V_1}{e^2\phi^*\phi}+\frac{i}{e\phi}\xi$. So we can now write \eqref{Magseg} as
\begin{equation}\label{Magdiag}
\begin{split}
\mathcal{L}_{mag} =&~V_1^*V_1-V_2^*V_2+\frac{\kappa}{4\pi}\left[\left(\frac{\tilde{f}_{xy}}2-\frac{f_{xy}}{e^2\phi^*\phi}\right)(V_1+V_1^*)+\left(\frac{\tilde{f}_{xy}}2+\frac{f_{xy}}{e^2\phi^*\phi}\right)(V_2+V_2^*)\right]\\
&+\frac{i\kappa f_{xy}}{4\pi e\phi}\xi-\frac{i\kappa f_{xy}}{4\pi e\phi^*}\xi^*+|D_i\phi|^2.
\end{split}
\end{equation}
This is quadratic in $V_1$ and $V_2$, but only linear in $\xi$, which acts as a Lagrange multiplier that imposes the constraint $f_{xy} = 0$. Immediately using this constraint, \eqref{Magdiag} further simplifies to
\begin{equation}
\begin{split}
\mathcal{L}_{mag} =&~V_1^*V_1-V_2^*V_2+\frac{\kappa\tilde{f}_{xy}}{8\pi}(V_1+V_1^*+V_2+V_2^*)+|D_i\phi|^2,\\
=&\left(V_1+\frac{\kappa\tilde{f}_{xy}}{8\pi}\right)^*\left(V_1+\frac{\kappa\tilde{f}_{xy}}{8\pi}\right)-\left(V_2-\frac{\kappa\tilde{f}_{xy}}{8\pi}\right)^*\left(V_2-\frac{\kappa\tilde{f}_{xy}}{8\pi}\right)+|D_i\phi|^2,
\end{split}
\end{equation}
where we have added and subtracted the same term to complete both perfect squares. Integrating out these perfect squares, we are left with
\begin{equation}\label{Duh}
\mathcal{L}_{1} = |D_i\phi|^2
\end{equation}
with the constraint $f_{xy} = 0$.

\subsection*{If $\xi$ is not integrated out}
If we keep $\xi$ as a field in the reduced theory, We go back to \eqref{Magdiag}. Now we don't impose $f_{xy} = 0$ since $\xi$ is no longer a Lagrange multiplier. We can again complete the squares involving $V_1$ and $V_2$, and doing that we get
\begin{equation}
\begin{split}
\mathcal{L}_{mag} =&~\left(V_1+\frac{\kappa\tilde{f}_{xy}}{8\pi}-\frac{\kappa f_{xy}}{4\pi e^2\phi^*\phi}\right)^*\left(V_1+\frac{\kappa\tilde{f}_{xy}}{8\pi}-\frac{\kappa f_{xy}}{4\pi e^2\phi^*\phi}\right)-\left(\frac{\kappa}{4\pi}\right)^2\left(\frac{\tilde{f}_{xy}}2-\frac{f_{xy}}{e^2\phi^*\phi}\right)^2\\
&-\left(V_2-\frac{\kappa\tilde{f}_{xy}}{8\pi}-\frac{\kappa f_{xy}}{4\pi e^2\phi^*\phi}\right)^*\left(V_2-\frac{\kappa\tilde{f}_{xy}}{8\pi}-\frac{\kappa f_{xy}}{4\pi e^2\phi^*\phi}\right)+\left(\frac{\kappa}{4\pi}\right)^2\left(\frac{\tilde{f}_{xy}}2+\frac{f_{xy}}{e^2\phi^*\phi}\right)^2\\
&+\frac{i\kappa f_{xy}}{4\pi e\phi}\xi-\frac{i\kappa f_{xy}}{4\pi e\phi^*}\xi^*+|D_i\phi|^2.
\end{split}
\end{equation}
Integrating out the perfect squares we get
\begin{equation}
\mathcal{L}_{mag} = 2\left(\frac{\kappa}{4\pi e}\right)^2 \frac{f_{xy}\tilde{f}_{xy}}{\phi^*\phi}+\frac{i\kappa f_{xy}}{4\pi e\phi}\xi-\frac{i\kappa f_{xy}}{4\pi e\phi^*}\xi^*+|D_i\phi|^2.
\end{equation}
If we change our minds now and integrate out $\xi$, it sets $f_{xy} = 0$ giving back the action \eqref{Duh}.

At this juncture, it is not clear to us what the null reduction of the magnetic CSM theory is hinting at. We hope to come back to this in more detail in the near future.

\bibliographystyle{JHEP}
\bibliography{ccft}
\end{document}